# Improving the energy-extraction efficiency of laser-plasma accelerator driven free-electron laser using transverse-gradient undulator with focusing optics and longitudinal tapering

G. Zhou,[1,2] Y. Jiao,[1,*] J. Wu,[3] T. Zhang[4]

[1] *Institute of High Energy Physics, Chinese Academy of Sciences, Beijing 100049, China*
[2] *University of Chinese Academy of Sciences, Beijing 100049, China*
[3] *SLAC National Accelerator Laboratory, Menlo Park, California 94025, USA*
[4] *Shanghai Institute of Applied Physics, Chinese Academy of Sciences, Shanghai 201800, China*

**Abstract**
It is reported that [Z. Huang *et al*., Phys. Rev. Lett. **109**, 204801 (2012)], high-gain free-electron laser (FEL) can be generated by transverse-dispersed electron beams from laser-plasma accelerators (LPAs) using transverse-gradient undulator (TGU) assuming an ideal constant dispersion function without focusing optics. The constant dispersion function keeps electrons beyond the resonant energy bandwidth still being on resonant with the FEL radiation. Instead, in this paper, the case with focusing optics in an LPA-driven FEL using TGU is numerically studied, in which the dispersion function should be monotonously decreasing along the undulator. Even though the FEL resonance is not always satisfied for off-energy electrons in this case, through subtly optimizing the initial dispersion and focusing parameters, it is feasible to achieve a similar radiation power to the case assuming an ideal constant dispersion function without focusing optics, and meanwhile, to attain a good transverse coherence. Moreover, higher radiation power could be expected if longitudinally tapering, transverse-gradient variation in different sections are systematically conducted, which is demonstrated with intense global multiparameters optimizations based on GENESIS simulation.

## I. INTRODUCTION

Accelerator-based light sources, such as storage rings and free-electron lasers (FELs), greatly benefit fundamental research in physics, chemistry, materials science, biology, and medicine by producing intense tunable radiation ranging from the infrared to hard x-ray region [1]. As contrasted with synchrotron radiation, FEL holds better properties in coherence, monochromaticity, power, duration, and so on. However, an FEL facility is usually large and costly, because of the long linear accelerator (LINAC) together with its shielding [2]. Thus, efforts have been made to miniaturize the FEL design for years. One promising way is to use a laser-plasma accelerator (LPA), instead of traditional LINAC, to drive a high-gain FEL [3-6].

Compared to traditional LINAC, LPAs have much higher accelerating field gradient, smaller size and cost less, but produce an electron beam with lager energy spread. At present, LPA can produce high energy (~ 1 GeV), high peak current (~ 10 kA), and low emittance (~ 0.1 $\mu$m) electron beam with a relatively large energy spread of about 1% experimentally [7, 8]. Such a large energy spread terribly interferes the FEL gain process, which hinders LPAs from driving a high-gain FEL.

The physics behind energy spread affecting FEL gain could be simply understood from the well-known FEL resonance condition, i.e. for FEL with planar undulators, the on-axis resonance relationship between radiation and electron beam is

$$\lambda_r = \frac{1+K_0^2/2}{2\gamma^2}\lambda_u, \qquad (1)$$

where $K_0=0.934\lambda_u$ [cm]$B$[T], $\lambda_u$ is the undulator period, $B$ is the peak field of the undulator, $\gamma$ is the electron beam energy in unit of the rest energy. Energy spread would lead to a violation of the above equation, which impedes the FEL gain progress. Typically, for a high-gain FEL, the relative energy spread should be much smaller than the dimensionless Piece parameter of FEL [9, 10]. Mathematically, that is

$$\sigma_\delta \ll \rho = \left[\frac{1}{16}\frac{I_0}{I_A}\frac{K_0^2[JJ]^2}{\gamma_0^3\sigma_x^2 k_u^2}\right]^{1/3}, \qquad (2)$$

where $I_0$ is the peak current of the electron beam, $I_A\sim 17$ kA is the Alfvén current, $\sigma_x$ is the rms transverse beam size, $k_u=2\pi/\lambda_u$, and $[JJ]=J_0(\xi)-J_1(\xi)$ for planar undulator, with $\xi=K_0^2/(4+2K_0^2)$.

Energy spread, on the level of 1%, should decrease FEL power to a rather lower level, which degrades the resolution of diffraction imaging experiments [11]. To overcome the impediment caused by electron beam energy spread in the FEL gain process, approaches, such as transverse-gradient undulator (TGU) [12] and decompression [13], have been proposed and studied in detail. Recent study on TGU for high-gain FEL driven by LPAs points out that electron beam with an appropriate dispersion cooperating with TGU would increase the output radiation power significantly, about two orders, more than with decompression [5]. And experimental attempts are under going to realize LPA driven FEL based on TGU (e.g. SIOM-FEL see Ref. [14]). Hence, we only discuss FEL using TGU hereafter.

In Ref. [12], TGU was proposed to reduce the sensitivity to the electron beam energy spread for FEL oscillators. It has been proved that this idea also suits for single-pass high-gain FEL in Ref. [5]. By canting the magnetic poles, a linear transverse dependence of undulator field can be generated, like

$$\frac{\Delta K}{K} = \alpha x, \qquad (3)$$

where $\alpha$ is the transverse gradient of the undulator. For an electron beam dispersed horizontally, we can get

$$x = \eta_0 \frac{\Delta\gamma}{\gamma} = \eta_0 \delta. \qquad (4)$$

Properly choosing the dispersion

$$\eta_0 = \frac{2+K_0^2}{\alpha K_0^2}, \qquad (5)$$

and keeping it constant along the TGU, the spread in electron beam's energy would be compensated. Then, the resonance condition would be satisfied for electrons with different energy.

Here in this paper, we take focusing optics into account in the FEL process of a seeded LPA-driven FEL using TGU. In Section II, we discuss the variation of the dispersion function in a

TGU due to focusing optics and its effects on the FEL gain process. Moreover, longitudinal tapering and transverse-gradient variation, as well as frequency detunning, are introduced briefly and the effectiveness of these approaches is proved by simulation results of some typical cases. In Section III, the energy-extraction efficiency of an FEL using TGU with focusing optics and different power improving approaches are optimized separately, and comparisons are made to further understand each factor's contribution to the enhancement of energy-extraction efficiency. It is worth mentioning that well-benchmarked three-dimensional FEL simulation code, GENESIS 1.3, has been fully utilized in this paper for numerical investigations [15].

## II. POWER IMPROVING APPROACHES FOR HIGH-GAIN FEL USING TGU WITH FOCUSING OPTICS

### A. High-gain FEL using TGU with focusing optics

Electron beam with finite transverse emittance would expand when traversing the undulators. Thus, proper focusing is essential to maintain the electron beam transverse size for effective FEL interaction [16, 17]. FODO cell is a kind of common external focusing structure, widely used in accelerator and FEL facilities. In this paper, FODO cells with alternately arranged focusing and defocusing quadrupoles are used throughout the numerical studies, in which the quadrupoles are of the same strength. Fig. 1 illustrates the layout of FODO cells along the horizontal axis. The variation of dispersion function through the TGU with focusing optics can be studied by standard accelerator matrix calculation

$$\begin{pmatrix} \eta_f \\ \eta'_f \\ 1 \end{pmatrix} = M \begin{pmatrix} \eta_i \\ \eta'_i \\ 1 \end{pmatrix}, \quad (6)$$

where $M$ is the transport matrix, $\eta_i$, $\eta_f$ represents the dispersion before and after components respectively and the prime indicates $\partial/\partial Z$. At the entrance of TGU, $\eta'$ is always set to 0. The transport matrix can be obtained by left multiplication of each component with the following formula,

$$M = M_n ... M_i \cdot M_{i-1} ... M_1, \quad (7)$$

where the subscript represents the component order along the beam moving direction [18].

Under the thin-lens approximation, and with straightforward derivation, dispersion function in the two drift sections of the first FODO cell can be calculated according to Eq. (6), which is shown as below

$$\eta(z) = \begin{cases} \eta_i(1 - \dfrac{Z}{f}), & 0 \le Z \le L \\ \eta_i[1 - \dfrac{L}{f} - \dfrac{L}{f^2}(Z - L)], & L \le Z \le 2L \end{cases} \quad (8)$$

where $L$ is the distance between two contiguous quadrupoles, $f$ represents for the focal length and $Z$ is the distance between the electron beam and the TGU entrance. Because of the periodic quadrupole structure, the dispersion function in the following FODO cells can be obtained similarly. According to Eq. (8), using such quadrupole structure, one can get a monotonously decreasing dispersion along the TGU with a positive *f*, and control the dispersion decreasing rate by choosing the values of *L* and *f*. To prove our matrix analysis about the dispersion variation is practical, simulation with GENESIS time-independent mode, based on parameters shown in Table I, was done. As shown in Fig. 1, due to the effects of the strong focusing, the beam dispersion would decrease along the TGU and the simulation result agrees with matrix calculation well.

TABLE I. Main parameters for an FEL using TGU with focusing optics.

| Parameter | Symbol | Value |
|---|---|---|
| Beam energy | $E$ | 500 MeV |
| Norm. transv. emittance | $\gamma_0\varepsilon_x$ | 0.1 $\mu$m |
| Peak current | $I_p$ | 5000 A |
| Rel. rms energy spread | $\sigma_\delta$ | 2% |
| Undulator period | $\lambda_u$ | 2 cm |
| Undulator parameter | $K$ | 1.93 |
| Distance between quadrupoles | $L$ | 2.5 m |
| Resonant wavelength | $\lambda_s$ | 30 nm |
| Transverse gradient | $\alpha$ | 43 m$^{-1}$ |

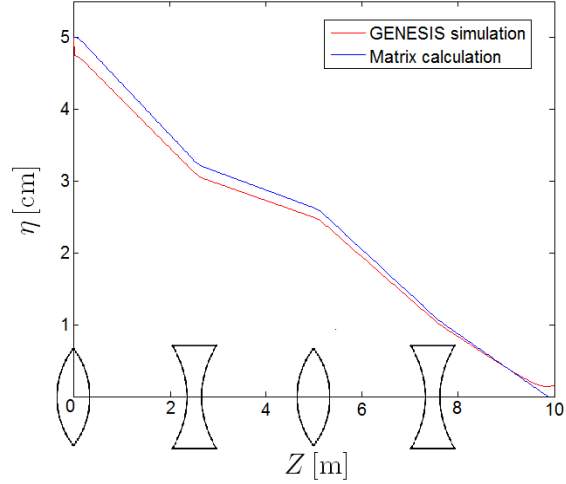

FIG. 1. Electron beam dispersion variation through a 10 m-TGU with initial dispersion equal to 5 cm and focal length equal to 7 m. Traditional lens symbols are used to represent for focusing quadrupoles and defocusing quadrupoles.

This decreasing dispersion leads to a reduction in transverse beam size, which improves the electron beam density [19]. This promotes the FEL gain process and reduces the transverse radiation field size. However, this effect does not contribute to the final FEL power gain significantly since the dispersion reduction also causes a mismatch between the transverse gradient and dispersion function. By optimizing the focusing parameters and the initial dispersion, a trade-off could be figured out, which is presented in Section III.

For a FODO cell we used, under the thin-lens approximation, the focusing parameters can be well described by the average $\beta$-function ($<\beta>$) and with straightforward derivation, one can get $<\beta>=2f$. In the following text, the authors would use $<\beta>$ to describe the focusing strength of FODO cells. FEL gain progress is affected by the dispersion function, which is controlled by $\eta_i$ and $<\beta>$. If other parameters are fixed, the radiation power is determined by $\eta_i$ and $<\beta>$. To find the relationship among $\eta_i$, $<\beta>$, and $P(\eta_i,<\beta>)_{max}$, very approximately, the authors assume that when $\eta_i$ and $<\beta>$ lead to the most efficient FEL gain progress, the mean of initial and final dispersion is about the dispersion, predicted by theoretical analysis [Eq. (5)], shown as below,

$$\eta_0 \sim \frac{\eta_i + \eta_f(\eta_i, <\beta>)}{2} \quad (9)$$

Assuming $\eta_0 = [\eta_i + \eta_f(\eta_i, <\beta>)]/2$, the above equation, is an implicit function about the $\eta_i$ and $<\beta>$. Taking five-meter, six-meter and seven-meter long TGU as examples, we plot the $\eta_i$ with respect to $<\beta>$ in Fig. 2. To verify this analysis, we did a grid scan over $<\beta>$ and $\eta_i$. For a fixed $<\beta>$, initial dispersion, achieving the highest radiation power, is plotted in Fig 2. By comparing the scatters and the curves, it can be found that this approximate analysis is quite feasible in the FEL exponential growth process. However, when the FEL gain process is about saturation, this rough analysis strays from the simulation result, like the green curve and points in the below figure. Based on Eq. (9) and Fig. 2, one can find that the initial dispersion should be larger than the theoretical dispersion because of the dispersion reduction along the TGU. Moreover, stronger external focusing or longer TGU would cause smaller final dispersion, as a result of which, the initial dispersion should be set larger to keep efficient FEL gain progress.

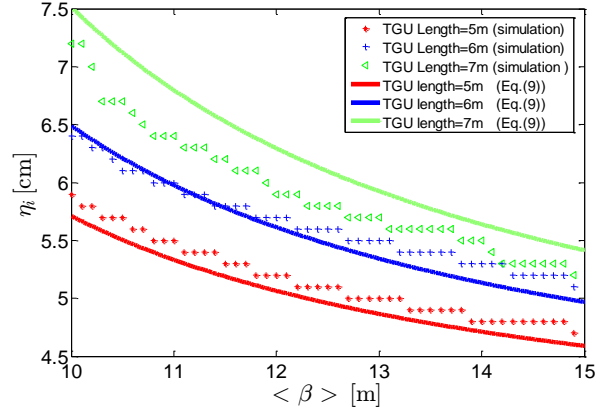

FIG. 2. For TGU with different length, the $\eta_i$ for fixed $<\beta>$ achieving the max $P(\eta_i,<\beta>)$ plotted as scatters and $\eta_i$-$<\beta>$ curve obtained from Eq. (9) plotted with solid lines.

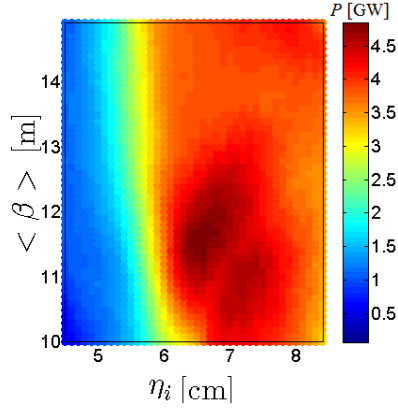

FIG. 3. Contour plot of the radiation power of an FEL using a ten-meter TGU with focusing optics with respect to the initial dispersion and average $\beta$-function.

Further, if the TGU length is larger than the saturation length, the relationship between $\eta_i$ and $<\beta>$ tends to be irregular, because of the energy exchange between the laser field and electron beam during the post-saturation progress [20]. This has been proved by grid scan over $\eta_i$ and $<\beta>$, supposing that the TGU length is 10 m, as shown in Fig. 3.

**B. FEL power improving approaches**

After FEL saturation, the radiation power oscillates around the saturation power [21]. Tapering can be used to further extract energy from the electron beam after FEL saturation through the TGU [22-24]. Here and after, we use a quadratic taper profile in our numerical study, which has been proved to be quite efficient. For comparison, GENESIS simulations of non-tapered TGU with focusing optics and tapered TGU with focusing optics have been done and the results are shown in Fig. 4. As shown in Fig. 4, we find that taper technology allows a significant improvement in radiation power, about 5 times in this case.

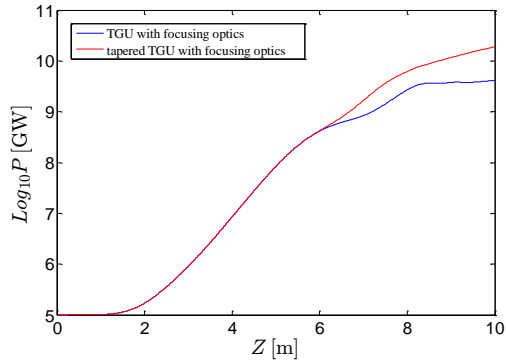

FIG. 4. GENESIS time-independent simulation of an FEL using a tapered TGU with focusing optics and a non-tapered TGU with focusing optics ($\eta$=7.5 cm, $<\beta>$=12.5 m), with taper staring from 5.5m and taper ratio equal to 0.07.

Considered the dispersion variation, the relationship between dispersion function and transverse gradient would be violated [see Eq. (5)]. To keep Eq. (5) satisfied along the TGU, the transverse gradient can be enlarged with dispersion function synchronously, which indicates that

$$\frac{2+K_0^2}{K_0^2} = \eta(Z)\alpha(Z). \qquad (10)$$

Since an undulator with gradually increasing transverse gradient is impracticable at present, we evenly separate the undulator into two sections with different transverse gradient to roughly study the power improvement brought by this remedy. The effectiveness of this idea is proved by GENESIS simulation as shown in Fig. 5.

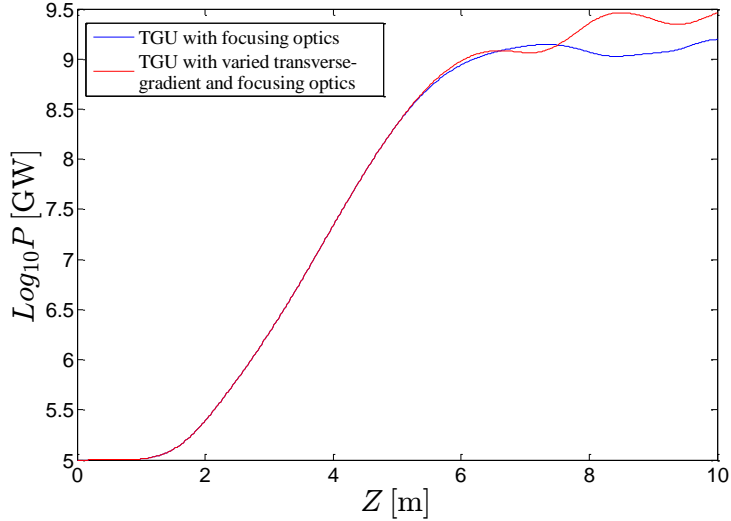

FIG. 5. Time-independent simulation of FEL using TGU with varied transverse-gradient and focusing optics contrasted with that without varied transverse-gradient. The initial dispersion is 5 cm and the average $\beta$-function is chose to be 14 m. The transverse gradient is 83 T/m for the 2$^{nd}$ TGU section and 30T/m for the 1$^{st}$ TGU section.

The frequency of which the FEL power grows faster is somewhat lower than that predicted by Eq. (1) [20]. And this also suits for FEL using TGU without focusing optics, proved in Ref. [25] by theoretical analysis and simulation. This indicates that frequency detuning may help to improve the radiation power in our case. To study the influence of frequency detune on the radiation power, simulations of FEL using TGU with focusing optics resonating on seed laser frequency and detuning frequency are done. Analogy to FEL using TGU without focusing optics, empirically, the detuning frequency differs a little from the resonance frequency. The simulation result is shown in Fig. 6, which proves that for a seeded-FEL, to ensure the frequency whose power grows fastest equal to the seed laser frequency, the undulator parameter $K$ should be a little bit smaller than $K_0$ obtained from Eq. (1).

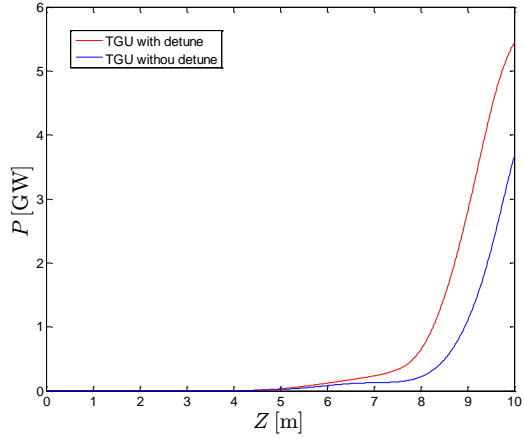

FIG. 6. Time-independent simulation results of an FEL using TGU with focusing optics resonating on the seed laser frequency and a detuning frequency, with initial dispersion and average $\beta$ function equal to 6.95 cm and 15 m. The red curve ($\Delta K/K_0$=-0.004) shows the benefit with detuning frequency.

## III. NUMERICAL OPTIMIZATION EXAMPLE OF AN LPA DRIVEN FEL USING TGU WITH FOCUSING OPTICS

In this section, we study the power optimization of a compact seeded-FEL driven by an LPA with taper, varied transverse-gradient, frequency detuning and so on. For physically realistic situations, we need to optimize the FEL radiation power within a finite TGU length. According to Fig. 2 and Fig. 3, considered compactness and radiation power improvement, we choose to optimize the radiation power of a compact EUV-FEL driven by LPAs using a 10-meter long TGU with focusing optics as radiator.

To find the optimal results within a shorter time, RCDS, a both efficient and robust optimization algorithm was used in this study. Here and after, optimizations we mentioned are done using this algorithm [26]. The power optimizations based on GENESIS time-dependent and time-independent simulation are done separately. Three simple models, (non-tapered) TGU with focusing optics, tapered TGU with focusing optics and tapered TGU with focusing optics and varied transverse gradient, are optimized to compare the different power improvement approaches' influence on optimized radiation power.

**A. Optimization for the steady-state cases**

The optimized radiation power of FEL using TGU with focusing optics and that without focusing optics are compared first to ensure that focusing optics does not lower the radiation power. For the case of non-tapered TGU with focusing optics, $\eta_i$, $<\beta>$ and $\Delta K/K_0$ (representing for frequency detuning effect) are used as optimizing variables. And for the case of TGU without focusing optics, $\eta_i$, rms horizontal beam size, rms vertical beam size, horizontal phase advance, vertical phase advance and $\Delta K/K_0$ are set as simulation variables to be optimized. The results are shown in Table II below. As can be seen in Table II, based on such a set of parameters, the optimal radiation power of an FEL using TGU with focusing optics is a little bit higher than that of TGU without focusing optics. And a simple study on tolerance of initial dispersion is done base on optimized parameters of TGU with/without focusing optics, shown in Fig. 7. One can find that radiation power of FEL using TGU without focusing optics is more sensitive to the deviation from the optimized initial dispersion.

TABLE II. Optimization result of TGU without focusing optics and with focusing optics.

| Parameters | Without focusing optics | With focusing optics |
|---|---|---|
| $\eta_i$ | 3.57 cm | 6.93 cm |
| $<\beta>$ | – | 12.94 m |
| $K_w/K_0$ | 0.9946 | 0.9953 |
| $P_{max}$ | 6.8 GW | 7.3 GW |

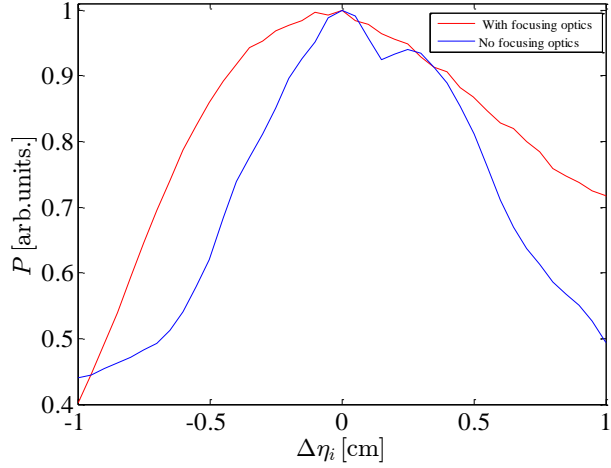

Fig. 7. The influence of initial dispersion deviation on the output radiation power of TGU with/without focusing optics.

To further improve the radiation power of FEL using TGU with focusing optics, as aforementioned, a quadratic taper part is added to the TGU to further extract energy from the electron beam after saturation, while the TGU length is remained the same. The added variables to be optimized, compared to non-tapered TGU with focusing optics, are the point taper starts ($Z^*$), taper ratio ($k$). The results are listed below in Table III together with that of non-tapered TGU with focusing optics for contrast.

As shown in Table III, by comparing the optimized radiation power of FEL using TGU with focusing optics and tapered TGU with focusing optics, taper provides a four times radiation power improvement. The optimal parameters of tapered TGU with focusing optics, compared to those of TGU with focusing optics, have smaller $\eta_i$, $\beta$, $K/K_0$. This leads to faster power growth before taper start point.

TABLE III. Optimization result of three simple TGU models without time-dependent effects

| Parameters | With focusing optics | With tapering and focusing optics | With tapering, varied transverse-gradient and external focusing |
| --- | --- | --- | --- |
| $\eta_i$ | 6.93 cm | 6.39 cm | 6.16 cm |
| $<\beta>$ | 12.94 m | 12.42 m | 11.21 m |
| $K_w/K_0$ | 0.9953 | 0.9980 | 0.9991 |
| $Z^*$ | - | 5.84 m | 5.63 m |
| $k$ | - | 0.085 | 0.102 |
| $K_w/K_0$ (2nd) | - | - | 0.9960 |
| $\Delta\alpha$ | - | - | 37.85 T/m |
| $\Delta Q$ | - | - | -0.97 |
| $P_{max}$ | 7.3 GW | 28.9 GW | 34.5 GW |

As shown in the above section, in some degree, transverse-gradient increment in the second section of the TGU improves the radiation power by remedying the mismatch between the dispersion function and the transverse gradient. For FEL using tapered TGU with focusing optics and varied transverse-gradient, additional parameters to be optimized compared with tapered TGU with focusing optics are the increment of transverse gradient ($\Delta\alpha$), the change of quadrupole strength ($\Delta Q$) and the detune parameter ($K_w/K_0$) in the 2nd section. The optimized parameters are also listed in Table III.

It can be seen that, the gradient increment in the second section of the TGU provides an additional radiation power about 20% as compared with that of tapered TGU with focusing optics. The optimal parameters, contrasted with those of tapered TGU with focusing optics, $\eta_i$, $\beta$, $K/K_0$ are smaller and taper starts earlier and taper ratio is larger. This indicates that transverse-gradient variation in different sections allows a more aggressive taper to extract energy from the electron beam.

**B. Optimization for the time-dependent cases**

To understand the influence of time-dependent effects on these three kinds of TGUs, optimizations on the radiation power obtained by the GENESIS time-dependent simulation are done. Note that, the fact that the radiation slips out of the electron beam is not considered here. The parameters to be optimized are the same as those of the power optimization based on GENESIS time-independent simulation and the optimizing objective turns to be the average radiation power over the electron beam. The initial beam current is set as a constant profile same as the case of time-independent to exclude the influence from the beam profile. The result is shown in Table IV and the transverse mode pattern of the radiation field at the optimized peak power of FEL using these TGUs are shown in Fig. 8. Since our study focuses on a seeded-FEL, the temporal coherence of the radiation power is quite well, which we do not discuss in this paper.

TABLE IV. Optimization result of three simple TGU models considered time-dependent effects, supposing a flattop beam distribution.

| Parameters | With focusing optics | With tapering and focusing optics | With tapering, varied transverse-gradient and external focusing |
|---|---|---|---|
| $\eta_i$ | 7.11 cm | 6.77 cm | 6.48 cm |
| $<\beta>$ | 11.35 m | 12.90 m | 11.46 m |
| $K_w/K_0$ | 0.9951 | 0.9983 | 0.9983 |
| $Z^*$ | - | 6.16 m | 6.42 m |
| $k$ | - | 0.070 | 0.076 |
| $K_w/K_0$ (2nd) | - | - | 0.9965 |
| $\Delta\alpha$ | - | - | 29.55 T/m |
| $\Delta Q$ | - | - | -0.96 |
| $P_{max}$ | 2.85 GW | 6.46 GW | 12.97 GW |

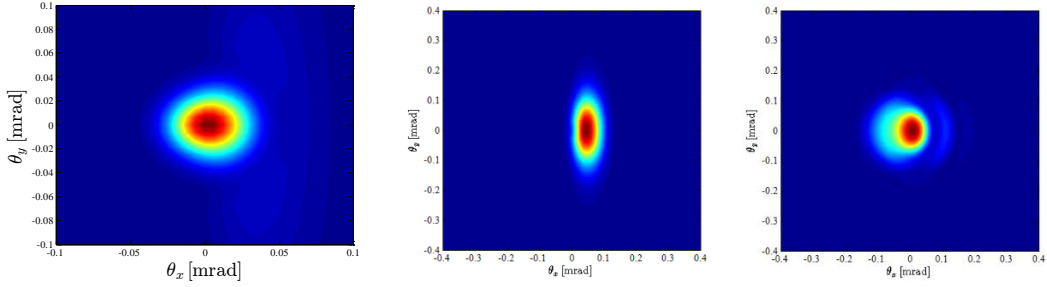

Fig. 8. Transverse mode pattern for FEL using TGU with focusing optics (left), tapered TGU with focusing optics (middle) and tapered TGU with varied transverse-gradient and focusing optics (right) at the optimized radiation power.

As shown in Table IV, the optimized radiation power is lower than that obtained by GENESIS time-independent simulation, because of the time-dependent effects. Different from the optimization results based on GENESIS time-independent simulation, taper improves the radiation power about two times and varied transverse-gradient provides additional two times radiation power. One can find that, the power improvement brought by taper is weakened when time-dependent effects are considered. Compared with parameters about taper obtained by optimization based on time-independent simulation, we find that $Z^*$ is larger and $k$ is lower. Namely, when time-dependent effects are considered, FEL using TGU with focusing optics requires a gentler taper. It's worth mentioning that, although the transverse coherence is not set as an optimizing objective, the transverse coherence of the optimized result of TGU with focusing optics is pretty good, which is shown in Fig. 8. And with power improving approaches, such as longitudinal tapering, varied transverse-gradient, the transverse coherence will be degraded in some degree. Further study on optimization of both the transverse coherence and radiation power is under way and results will be presented in a forthcoming paper.

### IV. CONCLUSION AND DISCUSSION
In this paper, we studied the radiation power improvement of an LPA driven FEL using TGU with focusing optics. A discussion about the behavior of the dispersion variation and its influence on the radiation power is presented. Optimization with GENESIS simulation shows that, FEL using TGU with focusing optics generates radiation field with pretty good transverse coherence and have a better tolerance of initial dispersion, while producing radiation power as high as the case without focusing optics. Both time-dependent and steady-state simulations show that approaches like tapering, frequency detuning and external focusing, as well as dispersion mismatch compensating contribute more or less to the FEL radiation power enhancement for TGU with focusing optics. And for both simulation cases, the radiation power of an FEL using a ten-meter TGU with focusing optics can be increased by more than four folds with these approaches. However, the radiation transverse coherence might degrade as the radiation power is improved. Fully optimization of both radiation power and transverse coherence should be the next work to be done. Note that in this paper, the configuration of focusing optics is simply assumed to be periodically arranged FODO cells, but this assumption is not a must and better solution may be found when more quadrupole combinations are considered. Finally, the analytical analysis of the FEL gain process along the TGU with focusing optics still remains as an open topic, which would absolutely help us to further understand the physics behind the TGU with focusing optics.

## ACKNOWLEDGMENTS

The first author is grateful to Zhe Duan and Daheng Ji for useful discussion and comments. This work was supported by National Natural Science Foundation of China (11475202, 11405187) and Youth Innovation Promotion Association of Chinese Academy of Sciences (No. 2015009)